\documentstyle[12pt,epsfig]{article}

\date{}

\begin{document}

\title{{\LARGE\sf The State(s) of Replica Symmetry Breaking: Mean
Field Theories vs.~Short-Ranged Spin Glasses}\thanks{An earlier version of this 
paper was posted as cond-mat/0105282 under the title,
``Replica Symmetry Breaking's New Clothes''.}}
\author{
{\bf C. M. Newman}\thanks{Partially supported by the 
National Science Foundation under grant DMS-01-02587.}\\
{\small \tt newman\,@\,cims.nyu.edu}\\
{\small \sl Courant Institute of Mathematical Sciences}\\
{\small \sl New York University}\\
{\small \sl New York, NY 10012, USA}
\and
{\bf D. L. Stein}\thanks{Partially supported by the 
National Science Foundation under grant DMS-01-02541.}\\
{\small \tt dls\,@\,physics.arizona.edu}\\
{\small \sl Depts.\ of Physics and Mathematics}\\
{\small \sl University of Arizona}\\
{\small \sl Tucson, AZ 85721, USA}
}

\maketitle

\begin{abstract}
We prove the impossibility of recent attempts to
decouple the Replica Symmetry Breaking (RSB) picture for finite-dimensional
spin glasses from the existence of many {\it thermodynamic\/} (i.e.,
infinite-volume) pure states while preserving another signature RSB feature
--- space filling relative domain walls between different finite-volume
states. Thus revisions of the notion of pure states cannot shield the
RSB picture from the internal contradictions that rule out its physical
correctness in finite dimensions at low temperature in large finite volume.
\end{abstract}


\small
\renewcommand{\baselinestretch}{1.25}
\normalsize

\section{Introduction}
\label{sec:intro}

In this paper we will describe what the mean-field picture and its central
component, replica symmetry breaking (RSB), {\it must\/} mean for
short-ranged spin glasses in all finite dimensions --- and why it {\it
cannot\/} hold for these systems.

In a recent paper$^{\footnotesize{\cite{MPRRZ}}}$ (hereafter [MPRRZ]),
Marinari, Parisi, Ricci-Tersenghi, Ruiz-Lorenzo, and Zuliani have provided
the most extensive description of the mean-field RSB picture offered to
date.  In response to earlier
demonstrations$^{\footnotesize{\cite{NS96a,NS96b,NSBerlin,NS97,NS98}}}$
(hereafter [NS96a], [NS96b], [NS97a], [NS97b], and [NS98], respectively) by
the authors that the mean-field RSB picture cannot describe the structure
of thermodynamic, i.e., infinite-volume, pure states at temperatures $T>0$
or ground states at $T=0$ of spin glasses in any finite dimension $d$,
[MPRRZ] proposed (see also Appendix 1 of the paper by Marinari
et~al.~$^{\footnotesize{\cite{MPR}}}$) that RSB is not meant to provide
such a description, but instead applies only to the structure of
``finite-volume pure states'', which are the relevant physical objects.  An
unambiguous definition of finite-volume pure states was not provided, but
it was emphasized that they were different from the ``pure states in an
infinite system'', i.e., the usual thermodynamic pure states.

However, in Sec.~\ref{sec:proof}, we present a new proof (whose
applications go well beyond spin glasses alone), which, when applied in the
current context, shows rigorously that the primary claims of
[MPRRZ]$^{\footnotesize{\cite{MPRRZ}}}$ are {\it incompatible\/} with each
other.  That is, if the claim of nontrivial link overlap ($P^L_e(q)$) for
large $L$ is valid, {\it it must give rise to multiple ground and pure
states in the usual thermodynamic sense\/} (cf.~Appendix A), as
traditionally envisioned (see, for example, the review
article$^{\footnotesize{\cite{BY}}}$ by Binder and Young, hereafter [BY],
or the book$^{\footnotesize{\cite{MPV}}}$ by Mezard, Parisi, and Virasoro,
hereafter [MPV]).  Therefore, whether or not a new interpretation of the
mean-field RSB theory in terms of ``finite-volume pure states'' can ever be
precisely formulated, the more usual infinite-volume interpretations cannot
be avoided.  But a mean-field structure for multiple {\it thermodynamic\/}
states has been ruled out by the authors' previous
arguments$^{\footnotesize{\cite{NS96a,NS96b,NSBerlin,NS97,NS98}}}$.
Mean-field RSB theory therefore cannot apply to realistic (i.e.,
finite-dimensional short-ranged) spin glasses.

The paper is organized as follows.  In Sec.~\ref{sec:review} we introduce
many of the terms and concepts needed for later sections, and provide an
abbreviated review of the issues concerning pure states within the
framework of mean-field RSB theory.  Sec.~\ref{sec:pinning} discusses the
behavior of interfaces, or domain walls, in finite volumes and how they can
(or cannot) give rise to multiple thermodynamic ground or pure states.
Sec.~\ref{sec:fvps} discusses the notion of ``finite-volume pure states''
introduced in [MPRRZ]$^{\footnotesize{\cite{MPRRZ}}}$, and provides an
initial critique of this concept, and Sec.~\ref{sec:links} reviews the
predictions of mean-field RSB theory for interface properties.  In
Sec.~\ref{sec:proof} we formally state our theorem that the mean-field RSB
theory in fact {\it must\/} predict multiple thermodynamic pure state pairs
with properties that have been previously ruled out; and in
Sec.~\ref{sec:discussion} we discuss the implications of our theorem and
present our conclusion.

We also include two appendices.  Appendix~A is a brief summary of the
definitions and properties of finite-volume Gibbs states, infinite-volume
Gibbs states, and pure states.  These play a major role in the text.
Appendix~B is a glossary providing brief definitions of other terms
frequently used in the text; some are in common usage in the literature,
but most are less so.

\section{Background and review}
\label{sec:review}

In this section we provide a review of the concepts and terms that will be
used throughout, and provide an abbreviated overview of recent developments
in the equilibrium theory of finite-dimensional spin glasses, as they
pertain to the current discussion. Detailed presentations can be found in
the references cited.  This section is included so that this paper is
reasonably self-contained; readers familiar with these topics may want to
skip ahead to Sec.~\ref{sec:pinning}.

For specificity, we focus on the
Edwards-Anderson$^{\footnotesize{\cite{EA}}}$ (EA) Ising spin glass, whose
Hamiltonian is given by
\begin{equation}
\label{eq:Hamiltonian}
{\cal H}=-\sum_{<x,y>} J_{xy} \sigma_x\sigma_y \ ,
\end{equation}
where the couplings $J_{xy}$ are independently chosen from a Gaussian
distribution with mean zero and variance one, the sum is over only nearest
neighbors on the $d$-dimensional cubic lattice ${\bf Z}^d$, and the spins
$\sigma_z=\pm 1$.  We note for later, however, that our results apply to a
wide range of models, including systems other than spin glasses.

\subsection{Essential features of the mean-field RSB picture}
\label{subsec:RSB}

Nontrivial replica symmetry breaking within the ``mean-field picture'' is
associated with a number of remarkable properties, including the existence
of many equilibrium states, non-self-averaging of overlap functions,
ultrametricity of pure state overlaps, and several others less relevant to
the current discussion. This picture is believed to describe the
low-temperature phase of the infinite-ranged
Sherrington-Kirkpatrick$^{\footnotesize{\cite{SK}}}$ model, where the sum
in the Hamiltonian Eq.~(\ref{eq:Hamiltonian}) now runs over {\it all\/}
pairs of spins, and the variance of the coupling distribution is rescaled
to provide a sensible thermodynamic limit.  We assume that the reader is
largely familiar with this picture, and refer her/him to
[BY]$^{\footnotesize{\cite{BY}}}$ or [MPV]$^{\footnotesize{\cite{MPV}}}$
for an extensive and detailed description.  Throughout this paper we will
refer to this picture and its variations as the mean-field picture, to
adhere to common usage in the literature; but it should be kept in mind
that it is based on the Parisi
solution$^{\footnotesize{\cite{P79,P83,MPSTV84a,MPSTV84b}}}$ of the
Sherrington-Kirkpatrick infinite-ranged model.

Numerous authors have asserted that mean-field-like RSB should describe the
broken symmetry of the low-temperature phase of more realistic
short-ranged, finite-dimensional spin glass models as well.  Its basic
features, for a fixed $T<T_c$, have been described in many places (see,
e.g., Refs.~\cite{MPRRZ,MPR,BY,MPV,MP00a,MP00b,CPPS,RBY,C92,P92,P93,MPRR})
and can be summarized as follows: (1) the existence of many equilibrium
states not related by any simple symmetry transformation, and whose number
grows without bound as system size $L\to\infty$; (2) for fixed coupling
realization ${\cal J}$, a nontrivial probability distribution $P_{\cal
J}(q)$, supported on countably many values $q_{\alpha \beta}$, for the spin
overlap between two different replicas; (3) non-self-averaging, i.e.,
${\cal J}$-dependence, of $P_{\cal J}(q)$, so that averaging $P_{\cal
J}(q)$ over all ${\cal J}$ yields a $P(q)$ supported on a continuum of
values between $\pm q_{EA}$, with nonzero weight at $q=0$ and
$\delta$-function spikes at $\pm q_{EA}$; (4) ultrametricity of the spin
overlaps $q_{\alpha \beta}$ among the equilibrium states; (5) nontrivial
edge overlap $P_e(q_e)$; for example, if one chooses at fixed ${\cal J}$ a
ground state from a cube with periodic boundary conditions, and a second
ground state from the same cube with antiperiodic boundary conditions, then
there would be a nonvanishing density (as $L\to\infty$) of bonds satisfied
in one but not the other ground state.

To arrive at these features, the mean-field RSB picture postulates, as in
[MPRRZ]$^{\footnotesize{\cite{MPRRZ}}}$, that at fixed $T$ the
finite-volume Gibbs state $\rho^L_{\cal J}$ in $\Lambda_L$, the cube of
side-length $L$ centered at the origin (we henceforth assume periodic
boundary conditions for specificity, but in fact our arguments will apply
to any boundary conditions chosen independently of ${\cal J}$), is
approximately a mixture of many pure states (we defer until later the
question of what this actually means):
\begin{equation}
\label{eq:sum}
\rho_{\cal J}^{(L)}\approx\sum_\alpha 
W_{{\cal J},L}^\alpha\rho_{\cal J}^\alpha 
\end{equation}
where $W_{{\cal J},L}^\alpha$ represents the Boltzmann weight in
$\rho_{\cal J}^{(L)}$ of pure state $\rho_{\cal J}^\alpha$.  The
finite-volume overlap distribution $P_{\cal J}^L(q)$ is approximately the
corresponding mixture of many $\delta$-functions:
\begin{equation}
\label{eq:PL}
P_{\cal J}^L(q)\approx\sum_{\alpha,\gamma}W_{{\cal J},L}^\alpha 
W_{{\cal J},L}^\gamma
\delta(q-q_{\cal J}^{\alpha\gamma})\quad ,
\end{equation}
where $q_{\cal J}^{\alpha\gamma}$ is the overlap between the 
states $\alpha$ and $\gamma$:
\begin{equation}
\label{eq:qab}
q_{\cal J}^{\alpha\gamma} \approx
|\Lambda_{L}|^{-1}\sum_{x\in\Lambda_{L}}\langle\sigma_x\rangle^\alpha
\langle\sigma_x\rangle^\gamma \quad ;
\end{equation}
here $|\Lambda_{L}|$ is the number of sites in $\Lambda_{L}$.

An edge, or link, overlap distribution function can be similarly
constructed$^{\footnotesize{\cite{MPRRZ,MP00a,MP00b,CPPS}}}$.  For
simplicity, consider a ground state pair $\pm\sigma^L$ in $\Lambda_L$ with
periodic boundary conditions, and a second ground state pair $\pm\sigma'^L$
obtained in $\Lambda_L$, e.g., with antiperiodic boundary conditions.
There will be a relative domain wall (or walls) between the pairs
$\pm\sigma^L$ and $\pm\sigma'^L$, consisting of the set of bonds $\langle
xy\rangle$ in the dual lattice satisfied in one and not the other ground
state pair; that is, they obey
\begin{equation}
\label{eq:dw}
\sigma_x^L\sigma_y^L=-{\sigma'}_x^L{\sigma'}_y^L\ .
\end{equation}
The link overlap between $\pm\sigma^L$ and $\pm\sigma'^L$ is
\begin{equation}
\label{eq:edge}
q_{{\cal J},e}^{\sigma^L\sigma'^L}= |E_L|^{-1} \sum_{{\langle
x,y\rangle}\in E_L}\sigma_x^L\sigma_y^L{\sigma'}_x^L{\sigma'}_y^L\
\end{equation}
which is equal to one when $\sigma^L=\pm\sigma'^L$ and smaller than one
otherwise.  Here $E_L$ is the edge set of, and $|E_L|$ is the number of
edges in, $\Lambda_L$.  The edge overlap distribution function is then
given by 

\begin{equation}
\label{eq:qeab}
P_{{\cal J},e}^L(q_e)\approx\sum_{\alpha,\gamma}W_{{\cal J},L}^\alpha W_{{\cal
J},L}^\gamma \delta(q-q_{{\cal J},e}^{\alpha\gamma})\quad .
\end{equation}

While this interesting picture at first seems reasonably clear, on closer
inspection there are numerous problems in interpretation when applied to
realistic models.  Much of [MPRRZ]$^{\footnotesize{\cite{MPRRZ}}}$ is
devoted to arriving at a definition of an ``equilibrium'' or ``pure'' state
within the mean-field RSB picture; but leaving that issue aside for now,
there are numerous other questions that could affect interpretation of
numerical measurements.  For example, by what procedure are states, or
replicas, chosen, and from what distribution?  In computing $P(q)$, what
does one mean by the ``infinite-volume limit''?  What is meant by
non-self-averaging when its presence or absence may depend on the sequence
of steps used to compute overlaps?  Until these questions are clarified, we
are forced to leave most of the above equations as approximate relations.

To illustrate, consider the situation at $T=0$.  For fixed ${\cal J}$ and
$\Lambda_L$ with periodic boundary conditions, there will be a single pair
of ground states $\pm\sigma^L$.  The overlap function will therefore be a
pair of $\delta$-functions at $\pm 1$ for all $L$, and so the limiting
$P(q)$ is that same pair of $\delta$-functions, independently of ${\cal
J}$.  Does this imply a single pair of ground states, as predicted by the
droplet/scaling picture of Macmillan$^{\footnotesize{\cite{Mac}}}$, Bray
and Moore$^{\footnotesize{\cite{BM85,BM87}}}$, Fisher and
Huse$^{\footnotesize{\cite{FH86,HF87a,FH87b,FH88}}}$, and others?  Not
necessarily, because if there were many ground state pairs then
$\pm\sigma^L$ would change chaotically with $L$, though for any single $L$
one would see only a single pair.  The presence or absence of this {\it
chaotic size dependence\/$^{\footnotesize{\cite{NS92}}}$} 
(hereafter [NS92]) is a reliable
test$^{\footnotesize{\cite{NS97,NS92}}}$ of whether there are,
respectively, many ground state pairs or only a single pair.  But if there
are many ground state pairs, can one construct $P(q)$ in order to see them?

\subsection{Standard interpretation of the RSB mean-field picture}
\label{subsec:standard}

The most straightforward, and natural, interpretation of the features of
the RSB mean-field picture described above is that the ``pure'' states in
Eq.~(\ref{eq:sum}) are the usual thermodynamic pure states, which are
easily and unambiguously defined for the EA model (see, e.g., Appendix~A of
this paper and
[NS97a,NS97b,NS98]$^{\footnotesize{\cite{NSBerlin,NS97,NS98}}}$).  In this
interpretation, for almost every fixed ${\cal J}$ there would be an
infinite number of these states.  The spin overlap distribution function
would be nontrivial in the sense described in the preceding section, and
would satisfy the properties of non-self-averaging and ultrametricity
(including at $T=0$).  The edge overlap distribution function would
similarly be nontrivial and non-self-averaging.  Procedures for
constructing overlap distributions are provided in
[NS96a]$^{\footnotesize{\cite{NS96a}}}$.

This interpretation has generally been the standard view (see, e.g.,
Refs.~\cite{BY,MPV,C92,P92,RTV,PRT00}), and is one way to answer the
questions posed in the previous section.  It allows us to replace the
approximate relation Eq.~(\ref{eq:sum}) with an equality
\begin{equation}
\label{eq:sumea}
\rho_{\cal J}(\sigma)=\sum_\alpha W_{\cal J}^\alpha\rho_{\cal J}^\alpha (\sigma)\ ,
\end{equation}
where $\rho_{\cal J}(\sigma)$ is an infinite volume mixed Gibbs state for a
particular coupling realization ${\cal J}$, the $\rho_{\cal J}^\alpha$ are
infinite-volume pure states for that ${\cal J}$, and the $W_{\cal
J}^\alpha$ their corresponding weights in $\rho_{\cal J}$.

The other equations in Section~\ref{subsec:RSB} are similarly replaced with
exact relations.  The overlap random variable becomes
\begin{equation}
\label{eq:overlapea}
Q=\lim_{L\to\infty}|\Lambda_L|^{-1}\sum_{x\in\Lambda_L}\sigma_x\sigma'_x
\end{equation}
where $\sigma$ and $\sigma'$ are chosen from the product distribution
$\rho_{\cal J}(\sigma)\rho_{\cal J}(\sigma')$.  If $\sigma$ is drawn from
$\rho_{\cal J}^\alpha$ and $\sigma'$ from $\rho_{\cal J}^\gamma$, then it
follows that the overlap is the constant
\begin{equation}
\label{eq:qabea}
q_{\cal J}^{\alpha\gamma} = 
\lim_{L\to\infty}|\Lambda_L|^{-1}\sum_{x\in\Lambda_L}\langle\sigma_x\rangle^\alpha
\langle\sigma_x\rangle^\gamma \quad .
\end{equation}
The probability distribution $P_{\cal J}(q)$ of $Q$ is therefore given by
\begin{equation}
\label{eq:PJ(q)ea}
P_{\cal J}(q)=\sum_{\alpha,\gamma}W_{\cal J}^\alpha W_{\cal J}^\gamma
\delta(q-q_{\cal J}^{\alpha\gamma})\quad .
\end{equation}
Edge overlap distribution functions are similarly defined.

However, it was rigorously shown in [NS96a]$^{\footnotesize{\cite{NS96a}}}$
that {\it this standard interpretation of the mean-field picture cannot
hold at any temperature in any finite dimension\/}, because the $P_{\cal
J}(q)$ of Eq.~(\ref{eq:PJ(q)ea}) must be self-averaging, i.e., the same for
almost every ${\cal J}$.  This also rules out the possibility of nontrivial
ultrametricity among the thermodynamic pure states.  We note that the
(nonrigorous) arguments of Parisi and
Ricci-Tersenghi$^{\footnotesize{\cite{PRT00}}}$, claiming to support
ultrametricity among {\it all\/} the equilibrium states of
finite-dimensional spin glasses, are in fact also consistent with {\it
trivial\/} ultrametricity, such as that displayed in the droplet/scaling
two-state picture or the many-state chaotic pairs picture
(cf.~Sec.~\ref{subsec:metastates} below).  One must therefore adopt an
unconventional interpretation of the mean-field RSB picture if there is to
be any hope of its application to realistic spin glasses.

\subsection{The nonstandard interpretation of the mean-field RSB picture}
\label{subsec:nonstandard}

The standard interpretation just described is a natural extrapolation to
large lengthscales of numerical simulations necessarily done on cubes
$\Lambda_L$ with relatively small $L$.  What is typically done numerically,
of course, is to generate (usually with periodic boundary conditions,
assumed here for specificity) finite-volume equilibrium Gibbs states (see
Appendix A) in $\Lambda_L$ and then to measure the overlap distribution for
fixed ${\cal J}$; then repeat the procedure for different ${\cal J}$'s and
compute the disorder-averaged $P_L(q)$.  Doing this for several different
$L$'s allows one to examine finite-size scaling and other properties of
overlap functions.

Numerically, one has no choice but to follow this or some similar
procedure; but in [NS96b]$^{\footnotesize{\cite{NS96b}}}$ it was shown that
evidence for RSB arising from this approach can correspond to more than one
thermodynamic picture.  The above procedure, if extrapolated to arbitrarily
large $L$, gives rise to a $P(q)=\lim_{L\to\infty}P_L(q)$ without any
explicit or prior construction of thermodynamic states.  (Another procedure
that {\it does\/} first construct states and then computes overlaps is
given in [NS96a]$^{\footnotesize{\cite{NS96a}}}$).

In these numerical computations, replica symmetry is of necessity broken
{\it before\/} the $L\to\infty$ limit is taken.
Guerra$^{\footnotesize{\cite{Guerra}}}$ has pointed out that changing the
order of these limits can be quite significant.  That this interchange of
limits$^{\footnotesize{\cite{NS96b}}}$ can lead to a new thermodynamic
picture of the spin glass phase does not seem to have been appreciated
prior to [NS96a,NS96b].

Based on these considerations, a new, nonstandard interpretation of the
mean-field RSB picture was described in detail in Sec.~VII of
[NS97b]$^{\footnotesize{\cite{NS97}}}$; we provide only a brief summary
here.  It is a maximal mean-field picture, preserving mean-field theory's
main features, as discussed in Sec.~\ref{subsec:RSB}, although in an
unusual way.  The most natural description of this nonstandard
interpretation is in terms of the {\it metastate\/}, described in the next
section, but in order to simplify the discussion we forego use of the
metastate here.

As a starting point, then, this interpretation would mean that in any large
$\Lambda_L$, the Gibbs state is an approximate decomposition over many {\it
thermodynamic\/} pure states:
\begin{equation}
\label{eq:sumfinite}
\rho_{{\cal J}}^{(L)}(\sigma)\approx\sum_
\alpha W_{\cal J}^{\alpha,L}\rho_{\cal J}^\alpha (\sigma)\ ,
\end{equation}
where a few states dominate the sum in any fixed $L$. The overlap
distribution in $\Lambda_L$ for fixed ${\cal J}$, given by
Eq.~(\ref{eq:PL}), is nontrivial: for any fixed, large $L$ it would be a
sum of several $\delta$-functions, and the locations of these
$\delta$-functions would satisfy ultrametricity increasingly accurately as
$L\to\infty$.  When averaged over ${\cal J}$ at fixed $L$, this
distribution would broaden into a continuum between two $\delta$-functions
at $\pm q_{EA}$. 

Eq.~(\ref{eq:sumfinite}) and the properties listed after it all hold
equally well for the standard interpretation of mean-field RSB described in
Sec.~\ref{subsec:standard}.  The difference between the standard and
nonstandard interpretations arises from their thermodynamics; the
straightforward extrapolation to infinite volumes characteristic of the
standard interpretation is absent in the nonstandard picture.  In the
latter case, the infinitely many thermodynamic pure states are grouped into
``families'' of mixed states (the $\Gamma$'s of
Sec.~\ref{subsec:metastates}), each of which {\it individually} has the
properties listed in the preceding paragraph.  The union of all of these
families, which loses these properties, comprises the thermodynamic
structure of the nonstandard interpretation.

The crucial conceptual point is that the resulting {\it ensemble\/} of
overlap distributions remains independent of ${\cal J}$.  So while overlap
distributions still do {\it not\/} depend on ${\cal J}$, one now replaces
the usual notion of non-self-averaging over ${\cal J}$'s with a nonstandard
one: that is, averaging over $L$'s for fixed ${\cal J}$.  It can be shown
that this picture must have uncountably many pure states and overlaps, so
that {\it ultrametricity would not hold in
general\/}$^{\footnotesize{\cite{NS96a}}}$ {\it among any three pure states
chosen at fixed\/} ${\cal J}$, unlike in the standard interpretation (see,
for example, the papers of Vincent et al.$^{\footnotesize{\cite{VHO}}}$ and
of Badoni et al.$^{\footnotesize{\cite{BCPR}}}$).  Instead, each large
$\Lambda_L$ would pick out a subset of these (one of the families discussed
above) that {\it do\/} satisfy ultrametricity.

Unfortunately for the mean-field approach, it can also be shown that this
picture cannot hold at any temperature in any dimension, as discussed in
the next section.

\subsection{Metastates, chaotic pairs, and the simplicity of $P(q)$}
\label{subsec:metastates}

To explain why even the nonstandard interpretation cannot be valid, we need
to introduce the concept of {\it metastate\/}, discussed in detail in
[NS96b,NS97a,NS97b,NS98]$^{\footnotesize{\cite{NS96b,NSBerlin,NS97,NS98}}}$.
(For some uses of this concept in mean field models, see the papers of
K\"ulske$^{\footnotesize{\cite{Ku97,Ku98}}}$, Bovier and
Gayrard$^{\footnotesize{\cite{BG98}}}$, and Bovier et
al.$^{\footnotesize{\cite{BEN99}}}$.)  Metastates enable us to relate the
observed behavior of a system in large but finite volumes with its
thermodynamic properties.  This relation is relatively straightforward for
systems with few pure states or for those whose states are related by
well-understood symmetry transformations, as typically occurs in
homogeneous systems.  Experience with these has mostly guided intuition in
the case of disordered systems.  However, one of our early results is that,
in the presence of many pure states not related by any clear-cut symmetry
transformations, the relation between the system's thermodynamic properties
and its behavior in large but finite volumes may be non-obvious.

This is primarily due to the following result of
[NS92]$^{\footnotesize{\cite{NS92}}}$: if a system has many,
non-symmetry-related, pure states, the sequence of finite-volume Gibbs
measures generated using coupling-independent boundary conditions will
generally {\it not\/} converge to a single limiting thermodynamic state as
$L\to\infty$.  This is the phenomenon of {\it chaotic size dependence\/},
mentioned in Sec.~\ref{subsec:RSB}.  In the metastate approach, rather than
avoid this problem, we exploit it by focusing on an {\it ensemble\/} of
(pure or mixed) thermodynamic states.  This approach, based on an analogy
to chaotic dynamical systems, allows the construction of a limiting
measure.  Hence the term {\it metastate\/} --- while a thermodynamic state
is a probability measure on infinite-volume spin configurations (see
Appendix A), this new limiting measure is one {\it on the thermodynamic
states themselves\/}.

This infinite-volume measure has a particular usefulness in the context of
finite volumes because it tells us the likelihood of appearance of any
specified {\it thermodynamic\/} state, pure or mixed, in a typical large
volume.  More precisely, it provides a probability measure for all possible
$n$-point correlation functions contained in a box (or ``window''),
centered at the origin, whose sides are sufficiently far from any of the
boundaries so that finite size or boundary effects do not appreciably
affect the result.  (We discuss this in more detail in
Sec.~\ref{subsec:windows}.)

There are several ways of constructing metastates.  In
[NS96b,NS97a,NS97b]$^{\footnotesize{\cite{NS96b,NSBerlin,NS97}}}$ we
introduced the empirical distribution approach.  This considers, at fixed
${\cal J}$, a sequence of volumes with coupling-independent boundary
conditions.  Each finite-volume Gibbs state $\rho_{\cal J}^{(L_1)},
\rho_{\cal J}^{(L_2)},\ldots,\rho_{\cal J}^{(L_N)}$ in the sequence is
given weight $N^{-1}$.  This allows us to construct a histogram of
finite-volume Gibbs states; it was shown in [NS96b, NS97a] that this
histogram converges to a probability measure $\kappa_{\cal J}$ on the
thermodynamic states as $N\to\infty$.  A finite-volume Gibbs state in a
particular (large) volume approximates (deep in its interior --- cf.~the
remarks in the preceding paragraph) some infinite-volume thermodynamic
state $\Gamma$ restricted to that volume.  The resulting metastate
$\kappa_{\cal J}$ therefore specifies the fraction of cube sizes that the
system spends in each different thermodynamic state $\Gamma$.  An
individual $\Gamma$ may be either pure or mixed, depending on the system
and the boundary conditions used.

The empirical distribution approach presented above was shown in
[NS96b,NS97a]$^{\footnotesize{\cite{NS96b,NSBerlin}}}$ to be equivalent to
an earlier construction of Aizenman and
Wehr$^{\footnotesize{\cite{AW90}}}$.  In this alternative approach, the
randomness of the couplings is used directly to generate an ensemble of
states.  It can be proved that the two approaches are basically equivalent,
in that there exists at least a ${\cal J}$-independent subsequence of
volumes along which both methods yield the {\it same\/} limiting
metastate$^{\footnotesize{\cite{NSBerlin,NS97}}}$.

The metastate approach is specifically designed to consider both finite and
infinite volumes together and to unify the two cases. In essence, the
metastate provides the probability, for a randomly chosen large $L$,
of various thermodynamic pure (or ground, at $T=0$) states appearing inside
any fixed $\Lambda_{L_0}$.

We return now to the nonviability of the nonstandard interpretation of the
mean-field RSB theory in realistic spin glass models.  Our claim is based
on a simple theorem, presented in [NS98]$^{\footnotesize{\cite{NS98}}}$,
with a powerful implication --- that (at fixed $d$ and $T$) the metastate
$\kappa_{\cal J}$ is {\it invariant\/} with respect to flip-related
boundary conditions, chosen independently of the couplings.  That is, the
metastate constructed using periodic boundary conditions on the
$\Lambda_L$'s is the same as that constructed using antiperiodic boundary
conditions.  Even if one were to choose two {\it arbitrary\/} sequences of
periodic and antiperiodic boundary conditions, the metastates would {\it
still\/} be identical.  The metastate, and its corresponding overlap
distributions, is therefore highly insensitive to boundary conditions.

This metastate invariance has profound consequences.  It means that the
frequency of appearance of various thermodynamic states in finite volumes
is {\it independent\/} of the choice of periodic or antiperiodic boundary
conditions.  Moreover, this same invariance property holds among any two
sequences of {\it fixed\/} boundary conditions; the fixed boundary
condition may even be allowed to vary arbitrarily along any single sequence
of volumes!  It follows that, with respect to changes of boundary
conditions, the metastate is highly robust.

If there were only a single thermodynamic state, such as paramagnetic, or a
single pair of states as in droplet/scaling, this would be expected.  But
can this result can be reconciled with the presence of {\it many\/}
thermodynamic states?

The answer is yes, but it puts severe constraints on the form of the
metastate and overlap distribution functions.  In light of this strong
invariance property, any metastate constructed via coupling-independent
boundary conditions should be able to support only a very simple structure,
effectively ruling out the nonstandard interpretation of the mean-field RSB
picture.

How can this invariance property be reconciled with the presence of many
non-symmetry-related pure or ground states?  The only plausible possibility
is that in any metastate constructed from coupling-independent boundary
conditions (periodic, antiperiodic, free, fixed, etc.), all pure
thermodynamic states are equally likely. That is, each of these metastates
should be supported {\it uniformly\/} in some appropriate sense (which can
be made precise only with detailed knowledge about the pure states --- see,
e.g., the discussion in Sec.~IV of [NS98]$^{\footnotesize{\cite{NS98}}}$)
on the pure state pairs in that metastate.  This is the only plausible way
in which all sorts of different boundary conditions could give rise to the
same pure state distribution.

Such a uniform distribution, though, is inconsistent with the features of
the nonstandard mean-field picture.  That picture requires a {\it
non\/}uniform distribution over the pure states (for further discussion,
see [NS98]$^{\footnotesize{\cite{NS98}}}$), as does {\it any\/} picture in
which a nonzero fraction of $\Gamma$'s consists of a nontrivial mixture of
pure state pairs.  There is only one many-state picture of which we are
aware that is consistent with this theorem.  This is the {\it chaotic
pairs\/} picture, introduced in [NS92]$^{\footnotesize{\cite{NS92}}}$ and
[NS96b]$^{\footnotesize{\cite{NS96b}}}$ and further developed in
[NS97a,NS97b,NS98]$^{\footnotesize{\cite{NSBerlin,NS97,NS98}}}$.

The chaotic pairs picture resembles the scaling/droplet picture in finite
volumes, but has a very different thermodynamic structure.  It has {\it
infinitely\/} many thermodynamic pure states, but, unlike any mean-field
picture, in each large volume with periodic boundary conditions one
``sees'' only one pair of pure states at a time.  That is, for large $L$,
one finds that
\begin{equation}
\label{eq:possfive}
\rho_{\cal J}^{(L)}\approx {1\over 2}\rho_{\cal J}^{\alpha_L}+{1\over 2}\rho_{\cal J}^{-\alpha_L}
\end{equation}
where $-\alpha$ refers to the global spin-flip of pure state $\alpha$.  So
each $L$ picks out a single pure state pair from the infinitely many
present.  If all $\Lambda_L$ have periodic boundary conditions, then the
chaotic pairs picture would exhibit chaotic size dependence, unlike the
droplet/scaling picture.  In other words, in the scaling/droplet picture,
the low-temperature, periodic boundary condition metastate is supported on
one thermodynamic mixed state $\Gamma$ consisting of a single pure state
pair, and this $\Gamma$ is seen in a fraction one of the $\Lambda_L$'s.  In
chaotic pairs, the metastate is dispersed over infinitely many $\Gamma$'s,
of the form $\Gamma = \Gamma^\alpha={1\over 2}\rho_{\cal J}^\alpha+{1\over
2}\rho_{\cal J}^{-\alpha}$; here, two different $\Lambda_L$'s will
typically see {\it different but single\/} pure state pairs.

The overlap distribution for each $\Gamma$, hence each $\Lambda_L$ (for $L$
large) is the same:
\begin{equation}
\label{eq:PG}
P_\Gamma = {1\over 2}\delta(q-q_{EA})+{1\over 2}\delta(q+q_{EA})\ .  
\end{equation}
So the disorder-averaged spin overlap function $P(q)$ {\it and\/} link
overlap function $P_e(q_e)$, when constructed by breaking the replica
symmetry {\it before\/} taking the thermodynamic limit, as in
Sec.~\ref{subsec:nonstandard}, must have the same, simple structure whether
there exists a single pair of thermodynamic pure states or infinitely many:
the spin overlap function $P(q)$ would be a pair of $\delta$-functions at
$\pm q_{EA}$, and the link overlap function $P_e(q_e)$ a single
$\delta$-function at some $q_e(T)$ (see below), in {\it either\/} case.

There {\it is\/} a difference in overlap functions in the two pictures,
however, if the thermodynamic limit is taken {\it before\/} replica
symmetry is broken, as in Sec.~\ref{subsec:standard}.  Here, as already
noted, $P_{\cal J}(q)$ and $P_{{\cal J},e}(q_e)$ are, at fixed $T$ and $d$,
the same for almost every ${\cal J}$, regardless of which of the two
pictures actually occurs.  In droplet/scaling, $P_{\cal J}(q)$ is again a
pair of $\delta$-functions at $\pm q_{EA}$, whereas the link overlap
function (computed in a box small compared to $\Lambda_L$ and far from the
boundaries --- cf.~Sec.~\ref{subsec:windows}) $P_{{\cal
J},e}(q_e)=\delta(q_e-1)$ at $T=0$ and presumably remains a single
$\delta$-function at all temperatures, though the $q_e$ value where the
spike occurs decreases due to thermal fluctuations as $T$ increases.  In
chaotic pairs, $P_{\cal J}(q)$ would now most likely equal $\delta(q)$: it
was proven by the authors$^{\footnotesize{\cite{NS99}}}$ that $P_{\cal
J}(q)=\delta(q)$ for the spin overlaps of $M$-spin-flip-stable metastable
states for any finite $M$, and, if there are infinitely many ground state
pairs, we expect the same to be true for ground states, i.e., for
$M=\infty$.  The form of the edge overlap function in the chaotic pairs
picture, when replica symmetry breaking occurs after taking $L\to\infty$,
is less clear; the contribution coming from relative interfaces between the
many pairs of pure states may well be a $\delta$-function, but unlike any
two-state picture, would be supported on a link overlap $q_e<1$ even at
$T=0$.  (A lengthier discussion of link overlap functions in given in
Sec.~\ref{sec:links}.)

\begin{figure}[t]
\vskip 0.5in
\centerline{\epsfig{file=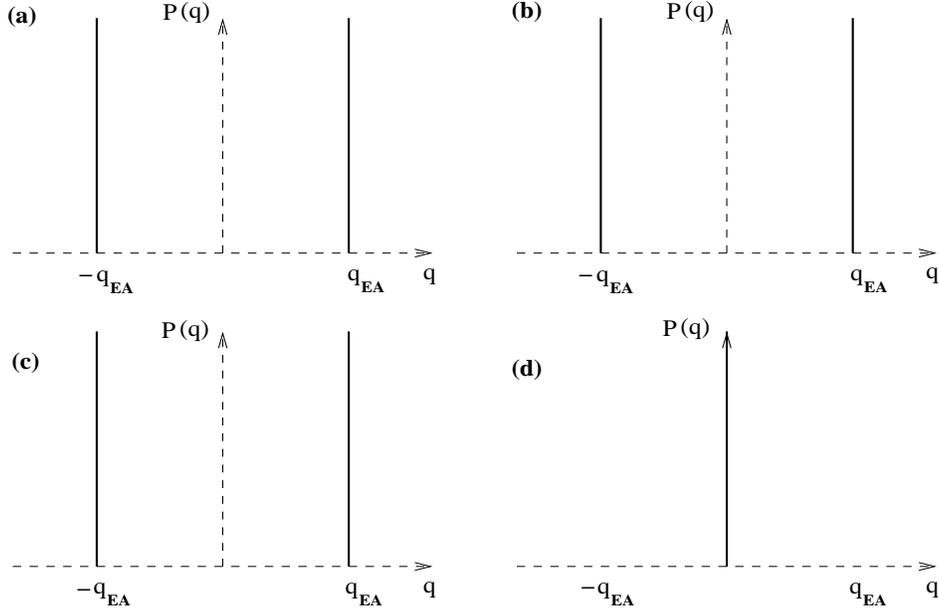,width=3.0in,height=2.8in}}
\vskip -0.35in
\renewcommand{\baselinestretch}{1.0} 
\small
\caption{The spin overlap function $P(q)$ at $T<T_c$ for: (a) a two-state
picture when replica symmetry is broken {\it before\/} taking the
thermodynamic limit; (b) the many-state chaotic pairs picture when replica
symmetry is broken before taking the thermodynamic limit; (c) a two-state
picture when replica symmetry is broken {\it after\/} taking the
thermodynamic limit; (d) the many-state chaotic pairs picture when replica
symmetry is broken after taking the thermodynamic limit (conjectured).}
\label{fig:overlap}
\end{figure}

Our conclusions are therefore that {\it the thermodynamic overlap structure
in spin glasses must be simple\/}, regardless of whether there are
infinitely many pure states or only a single pair.  The form of the overlap
function, however, can depend on how the computation is done.  Our results
for the spin overlap function $P(q)=P_{\cal J}(q)$ are summarized in
Fig.~\ref{fig:overlap}.

\renewcommand{\baselinestretch}{1.25}
\normalsize

In Fig.~\ref{fig:overlap}, the overlap function $P(q)$ is shown for two
very different physical pictures --- one a single pure state pair picture,
as in droplet/scaling, and the other the chaotic pairs picture, which
presupposes an uncountable infinity of pure states.  When comparing the
overlap function for different scenarios in general, it is important that
computations be done in the same way.  Figs.~1a and 1b represent overlap
computations done on cubes $\Lambda_L$ with periodic boundary conditions,
while Figs.~1c and 1d represent overlap computations done in infinite volume
on states randomly chosen from the respective periodic boundary condition
metastates.

The insensitivity of $P(q)$ (with all else remaining equal) to these very
different physical pictures indicates one potential problem with using
$P(q)$ for determining ground or pure state structure.  Figs.~1a and 1b are
identical because in either case a typical finite volume $\Lambda_L$
``contains'' only a single pure state pair.  If one instead looks at the
overlap of {\it all\/} of the infinite-volume pure states chosen (in this
example) from the periodic boundary condition metastate, as in Figs.~1c and
1d, the difference is evident.  In a two-state picture, one again sees a
single pair of delta-functions (Fig.~1c); the thermodynamic limit here is
straightforward because chaotic size dependence (periodic boundary
conditions again are assumed) is absent.  A many-state picture cannot have
the same $P(q)$ as the two-state picture when the replica symmetry is
broken {\it after\/} the thermodynamic limit is taken.  However, rather
than the nontrivial $P(q)$ one might expect, the invariance of the
metastate requires a very simple structure, as in Fig.~1d.

The forms of $P(q)$ sketched in Figs.~1c and 1d, however, are computed
using a procedure different from that used in numerical measurements, which
always use procedures corresponding to Figs.~1a and 1b.  Therefore the
usual measurements of $P(q)$ seem unable to provide unambiguous information
on pure state multiplicity or structure in realistic spin glasses.  (Other
numerical methods for distinguishing between two-state and many-state
pictures are described in [NS92,NS98]$^{\footnotesize{\cite{NS92,NS98}}}$
and in a more recent paper$^{\footnotesize{\cite{NS2D00}}}$ by us.)
If a measurement of $P(q)$ in a simple geometry (e.g.,
a cube) and with simple boundary conditions (e.g., free or periodic)
results in a complicated structure, it is likely that one is not
restricting the computation to a sufficiently small box far from the
boundaries (cf.~Sec.~\ref{subsec:windows}).

\subsection{Behavior at $T=0$}
\label{subsec:zero}

If the coupling distribution is continuous, such as Gaussian, then for any
finite $L$ and, say, periodic boundary conditions, there will be only a
single ground state pair $\pm\sigma^L$ in $\Lambda_L$.  If droplet/scaling
holds, then this pair will be the same (when restricted to a region far
from the boundaries --- see below) for all large $L$; if there are
infinitely many ground state pairs, then the pair changes chaotically with
$L$.  This will be true at $T=0$ for {\it any\/} many-state picture,
whether chaotic pairs, mean-field RSB, or some other such picture.  The
metastates, hence overlap functions, of these many-state pictures differ
only at positive temperature: the mean-field RSB picture at $T>0$ consists
of a nontrivial mixture of pure state pairs as in Eq.~(\ref{eq:sumfinite}),
while chaotic pairs looks similar at nonzero $T$ to its $T=0$ behavior.
That is, in chaotic pairs at $T>0$, the $\Gamma$ appearing in any
$\Lambda_L$ consists of a single pure state pair, as in
Eq.~(\ref{eq:possfive}).

The overlap distributions in Fig.~\ref{fig:overlap} should therefore apply
(either (a) and (c) or else (b) and (d), depending on whether
droplet/scaling or the chaotic pairs picture is correct) to both zero {\it
and\/} nonzero temperatures less than $T_c$.  The only temperature
dependence is in the magnitude of $q_{EA}$.

It is particularly important to note that {\it there is no difference
between the standard and nonstandard interpretations of the mean-field RSB
pictures at\/} $T=0$.  It follows that overlaps of ground states {\it
cannot\/} display nontrivial ultrametricity, or any other nontrivial
structure.  

Recent numerical results of Hed et~al.$^{\footnotesize{\cite{HHSD}}}$ have
claimed to see a nontrivial, hierarchical (though not ultrametric) ground
state structure for the $\pm J$ model in $3D$.  It is important to note
that the theorems described in previous sections apply to discrete coupling
models such as $\pm J$ as well as to continuous ones.  For all of these,
both the absence of non-self-averaging of overlap functions and the
invariance of the metastate are rigorous conclusions.  Overlap
distributions must therefore have a simple, or even trivial, form
regardless of the number of ground or pure states.  It seems likely,
therefore, that the results of Hed et~al.~are attributable to local
degeneracies that appear in the $\pm J$ model, rather than to any
nontrivial large-scale structures.

\subsection{Windows}
\label{subsec:windows}

Before we finish this review, we briefly mention that there remain
subtleties, alluded to in previous sections, in interpreting the results of
overlap measurements.  We refer the reader to the Appendix of
[NS97b]$^{\footnotesize{\cite{NS97}}}$ for a detailed discussion of the
effects of boundary conditions and of different methods of constructing
$P(q)$.  We also wish to emphasize a point discussed in detail in Sec.~VI
of [NS98]]$^{\footnotesize{\cite{NS98}}}$, where we discuss why, in order
to arrive at an accurate picture of the thermodynamic structure and the
nature of ordering of a system, one must focus attention on a fixed
``window'' near the origin.  A window (always defined in reference to the
volume $\Lambda_L$ under investigation) is a fixed cube $\Lambda_{L_0}$ in
$d$ dimensions, centered at the origin, and with $1\ll L_0\ll L$.  The
window lengthscale $L_0$ may be arbitrarily large, but must always be small
compared to the lengthscale $L$ of the entire volume $\Lambda_L$ under
consideration.  In particular, when examining the pure state structure of a
metastate, a window is simply any large cube centered at the origin with
fixed side $L_0\gg 1$.  This is because the metastate examines pure state
structure in a sequence of finite cubes $\Lambda_L$ with $L\to\infty$.
 
When calculating $P^L(q)$ and $P_e^L(q_e)$ in $\Lambda_L$, therefore, one
needs to do the overlap computation in a cube $\Lambda_{L_0}$ with $L_0\ll
L$, rather than in the entire volume as is usually done.  This condition is
difficult to achieve numerically, but cannot be avoided if one wants to
draw inferences about ordering of the low-temperature phase using overlap
functions.

This is not to say that computations done in the entire volume carry no
relevant or interesting information, only that their interpretation may be
unclear.  Such an example occurs in the numerical studies of Krzakala and
Martin$^{\footnotesize{\cite{KM00}}}$ (hereafter [KM]) and Palassini and
Young$^{\footnotesize{\cite{PY00}}}$ (hereafter [PY]).  These studies claim
to have uncovered a new type of excitation, which we have called {\it KMPY
excitations\/}$^{\footnotesize{\cite{NSregcong}}}$ (hereafter [NS01];
however, the numerical procedures used have been questioned --- see the
paper by Middleton$^{\footnotesize{\cite{Middleton01}}}$).  It was
rigorously shown in [NS01] that KMPY excitations do not yield new ground or
pure states, but, if they persist on large lengthscales, could be relevant
to the {\it excitation\/} spectrum in finite volumes.

\section{Pinning vs.~deflection of interfaces, and thermodynamic states}
\label{sec:pinning}

To pursue further the above idea, and also in preparation for the next
section, we briefly review a basic physical feature that distinguishes
thermodynamic pure states from (putative) non-thermodynamic ones.  The
discussion here will closely follow that of
[NS01]$^{\footnotesize{\cite{NSregcong}}}$; see also Sec.~VI of
[NS98]$^{\footnotesize{\cite{NS98}}}$.

To simplify the discussion, we focus on ground states; these are the
thermodynamic pure states at $T=0$.  The discussion can be extended to
$T>0$ pure states by considering interfaces, equivalently domain walls,
between two spin configurations chosen from different pure states.

Suppose one considers the finite-volume GSP $\pm\sigma_P^L$ corresponding
to a cube $\Lambda_L$ with periodic boundary conditions and $L$ large.  If
one then switches to antiperiodic boundary conditions, a new GSP
$\pm\sigma_{AP}^L$ is generated.  The two ground state pairs will have one or more
relative interfaces, or domain walls, consisting of the set of bonds
$\langle x,y\rangle$ (in the dual lattice) satisfying Eq.~(\ref{eq:dw}).
This finite-size domain wall consists of bonds that are satisfied in one
but not the other GSP; it is the boundary of the set of spins that are
flipped in going from $\pm\sigma_P^L$ to $\pm\sigma_{AP}^L$.

The question then arises: how could one know in principle whether there
exists more than one {\it thermodynamic\/} GSP?  These are infinite-volume
spin configurations whose energy cannot be lowered by the flip of any {\it
finite\/} subset of spins, and are generated by any convergent sequence of
finite-volume ground state pairs, such as $\pm\sigma^L$ with $L\to\infty$
(see Appendix A).

The answer is that if the domain wall between $\pm\sigma^L$ and
$\pm\sigma'^L$ is {\it pinned\/}, then there are multiple ground state
pairs.  By pinning we mean the following.  Consider a fixed window of size
$L_0$, which though finite can be arbitrarily large.  Apply the procedure
of generating ground state pairs $\pm\sigma_P^L$ by using periodic boundary
conditions on cubes $\Lambda_L$, with $L\gg L_0$, and ground state pairs
$\pm\sigma_{AP}^L$ generated with antiperiodic boundary conditions on the
same cubes. Observe $\pm\sigma_P^{(L,L_0)}$ and $\pm\sigma_{AP}^{(L,L_0)}$,
which are the two ground state pairs restricted to $\Lambda_{L_0}$.  If
their relative interface remains inside $\Lambda_{L_0}$ as $L\to\infty$ ,
then the interface is {\it pinned\/}. If there are many ground state pairs,
then the interface would converge, along different subsequences of $L$'s,
to different well-defined limits inside $\Lambda_{L_0}$.

These pinned domain walls are interfaces between true thermodynamic ground
state pairs.  This follows because the corresponding spin configurations
are limits of finite-volume ground state pairs (see Appendix~A).  However,
another method of constructing interfaces uses a single boundary condition
(typically periodic) and adds a perturbation, either by forcing a pair of
spins to take an opposite relative orientation from that in the ground
state, as in [KM]$^{\footnotesize{\cite{KM00}}}$, or by adding a bulk
perturbation to the Hamiltonian, as in [PY]$^{\footnotesize{\cite{PY00}}}$;
the two methods are believed to give equivalent results.  Consider, e.g.,
the method of Krzakala-Martin.  If the interface is {\it pinned\/}, then
one can prove again that it separates true thermodynamic ground state
pairs, as follows.  As in [NS01]$^{\footnotesize{\cite{NSregcong}}}$, let
the two spins be chosen randomly for each $\Lambda_L$ from the uniform
distribution on its sites.  Because in the $L\to\infty$ limit the two sites
will, with probability one, be outside any fixed window, the spin
configuration inside the window will have minimum energy (given its
configuration on the boundary of the window).  This proves the desired
result, because the resulting infinite-volume spin configuration cannot
have its energy lowered by flipping any finite subset of spins (which would
necessarily be inside some fixed window).

Pinning of interfaces by quenched disorder occurs in disordered
ferromagnets$^{\footnotesize{\cite{HH85,K85,BK94}}}$ for sufficiently large
$d$; but these interfaces have lower dimension than the embedding space.
One interesting feature of RSB is the prediction of interfaces with
dimension $d_s$ {\it equal\/} to that of the embedding space; this will be
discussed in more detail in the following sections.

\begin{figure}[t]
\vskip -3in
\centerline{\epsfig{file=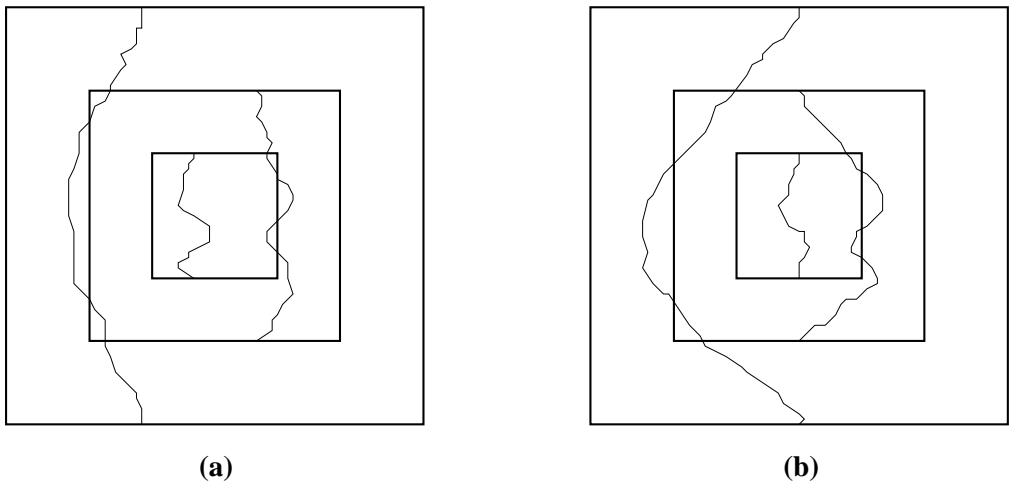,width=6.0in}}
\vskip -2.5in
\renewcommand{\baselinestretch}{1.0} 
\small
\caption{A sketch of interface deflection to infinity for a $2D$
disordered ferromagnet under (a) a change from periodic to antiperiodic
boundary conditions, and (b) a change from uniform plus boundary conditions
to Dobrushin boundary conditions (i.e., plus spins on the left
half-boundary of each square and minus on the right).  As $L$ increases,
the interface recedes from the origin in each case.  The interfaces
eventually are completely outside any fixed square.  (The deflection can
scale more slowly with $L$ than in the figure.)}
\label{fig:deflect}
\end{figure}

On the other hand, if the interface is {\it not\/} pinned, we say it
``deflects to infinity''.  Here, for any fixed $L_0$, the interface, for
all $L$ above some $L'$, will be {\it outside\/} $\Lambda_{L_0}$.  This is
what occurs with interface ground states in disordered
ferromagnets$^{\footnotesize{\cite{HH85,K85,BK94}}}$ for small $d$; see
Fig.~\ref{fig:deflect} for a schematic illustration.  If an interface
deflects to infinity, then it does not give rise to new thermodynamic pure
or ground states.

\renewcommand{\baselinestretch}{1.25}
\normalsize

\section{Finite-volume pure states --- replica symmetry breaking's new clothes}
\label{sec:fvps}

The various interpretations in Sec.~\ref{sec:review} of what the mean-field
RSB theory could mean for realistic spin glasses all used the usual concept
of ``pure state'' in its well-defined, traditional thermodynamic sense (see
Appendix~A).  However, in [MPRRZ]$^{\footnotesize{\cite{MPRRZ}}}$ it was
asserted that the ``pure states'' that have played a central role over the
past 20 years in the physical interpretation of the Parisi replica symmetry
breaking scheme$^{\footnotesize{\cite{MPV,P79,P83,MPSTV84a,MPSTV84b}}}$
should {\it not\/} be thought of in this way; instead, the relevant
physical objects are ``finite-volume pure states'' (not to be confused with
the usual finite-volume {\it Gibbs} states, as discussed in Appendix~A.)
Because of the potential importance of this re-interpretation of the
meaning of RSB in terms of finite-volume pure states, we now review (and
critique) this claim in this section.  The theorem and proof that in fact
RSB {\it must\/} involve the more traditional {\it thermodynamic\/} pure
states, which is the central result of this paper, will be presented in
subsequent sections.

The main new theoretical idea of [MPRRZ], introduced and
discussed in its Sec.~3, is the attempt to formalize the relation between
RSB and state structure for short-ranged models via the notion of {\it
finite volume pure states\/}.  This interpretation is contrasted with the
``not appropriate use of Eq.~(35) to describe an infinite system.''
Equation (35) of [MPRRZ] is simply:
\begin{equation}
\label{eq:inappropriate}
\langle\cdot\rangle = \sum_\alpha w_\alpha \langle\cdot\rangle_\alpha \, ,
\end{equation}
where $\alpha$ is a ``pure state'' index and $w_\alpha$ its Boltzmann
weight. 

Such a decomposition of course {\it can\/} be done and is well-defined
(see, e.g., the book of Georgii$^{\footnotesize{\cite{Georgii}}}$) for the
usual thermodynamic pure states $\alpha$ in infinite volume.  It can also
be done in a well-defined sense for {\it finite\/} volumes and is closely
related to the idea of ``window overlaps''; these ideas are introduced and
discussed in [NS98]$^{\footnotesize{\cite{NS98}}}$.  In both cases, our
theorems$^{\footnotesize{\cite{NS96a,NS96b,NSBerlin,NS97,NS98}}}$ apply and
rule out any of the interpretations in Sec.~\ref{sec:review} of the RSB
mean-field picture in finite-dimensional systems.  But a central point of
[MPRRZ] is that in RSB theory the decomposition
Eq.~(\ref{eq:inappropriate}) of a finite-volume Gibbs state does {\it
not\/} involve these traditional pure states, but rather a decomposition
into something else that does not have an infinite-volume definition or
meaning.

This is problematic in that it not only contradicts earlier statements of
the same authors, but also conflicts with other sections within the same
paper.  As to the former, references to pure states within the mean-field
RSB scheme that imply the usual thermodynamic definition are too numerous
to list.  As just one example, the RSB literature, whether discussing
infinite-ranged or finite-ranged spin glasses, repeatedly refers to pure
states as those having the clustering property given in
Eq.~(\ref{eq:clustering}) (see, for example,
[P83]$^{\footnotesize{\cite{P83}}}$, Sec.~III.F of [BY]$^{\footnotesize{\cite{BY}}}$, Sec.~3.1
in [MPV]$^{\footnotesize{\cite{MPV}}}$, [P92]$^{\footnotesize{\cite{P92}}}$); but clustering
appears to be an exclusive property of thermodynamic pure states.  Much of
the other terminology frequently used in the literature, such as
``valleys'', is vaguer.  Nevertheless, it is hard to interpret the
often-made claim that RSB pure states are separated by infinitely high
barriers (see, e.g., Sec.~IV.E of [BY] and Sec.~7.1 (and 3.1) of [MPV]), or
the dynamical assertion that a spin glass in one of these states would
thereafter spend an infinite amount of time in that state (see, e.g.,
Sec.~IV.E of [BY] and Sec.~7.1 (and 3.1) of [MPV]), as referring to
anything other than thermodynamic pure states.

It is implied in Sec.~3.1 of [MPRRZ] that finite volume pure states first
appeared in a 1987 paper of Parisi$^{\footnotesize{\cite{P87}}}$ (hereafter
[P87]), more than a decade earlier than [MPRRZ].  We believe that this is
unjustified.  There does not appear to be any discussion about finite
volume pure states in [P87], but rather discussion about ``pure clustering
states''.  As already noted, clustering is a property that belongs to
standard thermodynamic pure states.  Moreover, in [P87] the pure state
decomposition Eq.~(\ref{eq:inappropriate}) is justified several times on
the basis of a theorem in Ruelle's book$^{\footnotesize{\cite{Ruelle69}}}$
that is explicitly about thermodynamic pure states.

Even more seriously, there is a direct contradiction between the claim that
RSB refers only to ``nonthermodynamic'' pure states and the discussions in
Secs.~2.2, 2.3, 8.4, and 8.5 of [MPRRZ], where window overlaps
(cf.~Sec.~\ref{subsec:windows}) are discussed (see also a paper of Marinari
et~al.$^{\footnotesize{\cite{MPRR98}}}$).  Eqs.~(24) -- (27) of [MPRRZ]
concern the predictions of RSB for the spin overlap distribution confined
to a small region in the center of the cube; i.e., a window.  But these
predictions are precisely those that would be made by either the standard
or the nonstandard interpretation of the mean-field RSB picture described
in Secs.~\ref{subsec:standard} and \ref{subsec:nonstandard}; window
overlaps were especially constructed [NS98]$^{\footnotesize{\cite{NS98}}}$
so as to separate properties arising from {\it thermodynamic\/} pure state
structure from those due to boundary or other effects.

The final sentence of [MPRRZ] asserts that ``the recent rigorous results by
Newman and Stein strongly support RSB''.  Given the discussion in this
section and the earlier demonstration that our results imply that the {\it
only\/} sensible many-state picture is chaotic pairs, it should be clear
that we strongly dispute such a claim.  Its basis is discussed in Secs.~7.4
and 7.5 of [MPRRZ] (some of which appeared earlier as an unpublished
posting of Parisi$^{\footnotesize{\cite{Parcomment}}}$).  Our discussion
about why the arguments used in Ref.~\cite{Parcomment} (and [MPRRZ]) do not
support this assertion is provided in our own unpublished 
posting$^{\footnotesize{\cite{NSreply}}}$, 
to which we refer the reader.

Indeed, we will present below a rigorous result that, regardless of any new
intepretation of RSB based on finite volume pure states, excludes the
possibility that the mean-field RSB theory describes the low-temperature
spin glass phase in any finite dimension.

\section{RSB predictions for link overlaps and domain walls}
\label{sec:links}

We turn now to the central argument of our paper.  An unambiguous
prediction of the mean-field RSB approach, applied to finite-dimensional
spin glasses, is that the edge overlap distribution function $P_e(q_e)$
(Eq.~(\ref{eq:qeab})) is nontrivial on all length scales.  Extensive
discussion of the predictions of the RSB theory for realistic spin glasses
is given in several places; see in particular
[MPRRZ]$^{\footnotesize{\cite{MPRRZ}}}$,
[PY]$^{\footnotesize{\cite{PY00}}}$,
[MP00a]$^{\footnotesize{\cite{MP00a}}}$ and
[MP00b]$^{\footnotesize{\cite{MP00b}}}$, to which we refer the reader for
details.

It was noted in [MP00a,MP00b] that nontriviality in $P_e(q_e)$ at $T=0$
can be ascribed to the presence of space-filling domain walls between
ground states generated from different boundary conditions, or between
ground and excited states with the latter generated through a perturbation
(cf.~Sec.~\ref{sec:pinning}).  This important feature of RSB theory can be
described in the following way.

Consider a $d$-dimensional cube $\Lambda_L$, centered at the origin and
with periodic boundary conditions, so that for a given coupling realization
${\cal J}^L$ inside $\Lambda_L$ there exists a ground state pair
$\pm\sigma^L$.  Consider as before the spin configuration generated by
forcing a random pair of spins to take on an opposite orientation from that
in $\pm\sigma^L$, and then letting the resulting configuration relax to a
new state ${\sigma'}^L$ with minimum energy subject to this constraint
(alternatively, one could add a bulk perturbation as in [PY,MP00b]. Then a
central {\it physical\/} feature of the mean-field RSB picture
([MPRRZ,MP00a,MP00b]) is that $\pm\sigma^L$ and $\pm{\sigma'}^L$ differ in
the following ways: 1) their difference is global, i.e., there are $O(L^d)$
spins flipped in going from $\pm\sigma^L$ to $\pm{\sigma'}^L$; 2) the
lengthscale $l$ of their relative interface is $O(L)$, and the number of
bonds in the interface scales as $L^{d_s}$ with $d_s=d$, i.e., the
interface is space-filling; and 3) the energy of the relative interface
remains of order one independently of $l=O(L)$ so that the domain wall
energy scales as $l^{\theta'}$ with $\theta'=0$.  Domain walls with these
properties will henceforth be called {\it RSB interfaces\/}. It is easy to
see that, at $T=0$, properties (1) and (2) already give rise to nontrivial
$P_e(q_e)$ (and conversely that nontrivial $P_e(q_e)$ implies the existence
of interfaces with those properties).

What about $T>0$?  Now, because RSB asserts that each individual
low-temperature Gibbs state is a mixture of several states, it predicts
that even without any perturbations, there will be nontrivial $P(q)$ and
$P_e(q_e)$ inside $\Lambda_L$ as described in Sec.~\ref{subsec:RSB}.  This
prediction relies heavily on property (3) of the RSB interfaces, namely
that their energies are $O(1)$; if properties (1) and (2) were valid, but
not (3), this would lead to the chaotic pairs picture
(cf.~Sec.~\ref{subsec:metastates}) with many states but trivial overlap
distribution.

We note, however, that there is now a problem in interpretation, especially
for $P_e(q_e)$, because one needs to disentangle effects due to potential
multiple states from those due to normal thermal fluctuations.  One way of
doing this was discussed in Sec.~V of
[NS92]$^{\footnotesize{\cite{NS92}}}$; here one looks at two different cube
sizes and uses the presence or absence of chaotic size dependence to
differentiate between the two effects.  We propose another way here.  It is
known that, if the probability density function of the couplings is bounded
by a constant $C$, as in the usual Gaussian coupling case,
then$^{\footnotesize{\cite{NS92}}}$
\begin{equation}
\label{eq:bound}
1-\overline{\langle\sigma_x\sigma_y\rangle^2}\le 2Ck_BT
\end{equation}
in a cube $\Lambda_L$ with coupling-independent boundary conditions, such
as periodic.  Here an overbar represents an average over coupling
realizations.  
This bound is rigorously obeyed by a Gibbs state
generated from a {\it single\/} boundary condition, regardless of how many
pure states it contains.

But now suppose that one generates {\it two\/} Gibbs states at $T>0$ in
$\Lambda_L$, e.g., one with and one without a Palassini-Young bulk
perturbation ([PY]), as in [MP00b].  Then it should still be true that the
contribution to $P_e(q_e)$ from trivial thermal excitations would remain
bounded by $O(T)$, but the contribution from multiple RSB-like states, if
present, would not obey any such bound.  Therefore, at sufficiently low
$T$, thermal contributions to $P_e(q_e)$ should be negligible compared to
putative RSB contributions.  (As a consequence, we suggest that results
obtained at higher temperatures, like $0.7\, T_c$ as in some of the
simulations in [MPRRZ], are not useful in verifying the applicability of
RSB theory to realistic spin glasses.)

We now address the central question of this paper: are RSB
interfaces, which comprise a central feature of mean-field RSB theory,
compatible with the claims of [MPRRZ] that a thermodynamic
interpretation of RSB pure states can be avoided?  In other words, can RSB
domain walls {\it avoid\/} giving rise to many traditional thermodynamic
pure states?

We will provide a proof in the next section that the answer is {\it no\/};
these central predictions of mean-field RSB theory are {\it rigorously
incompatible\/} with each other.  The prediction of RSB interfaces means
that multiple {\it thermodynamic\/} pure states, with properties that have
been ruled out in our previous
papers$^{\footnotesize{\cite{NS96a,NS96b,NSBerlin,NS97,NS98}}}$, must
appear. The mean-field RSB theory is therefore {\it inconsistent\/} in any
finite dimension.

Before we turn to our rigorous result, we provide a heuristic argument that
illustrates the central idea of our theorem and makes clear why RSB
interfaces must give rise to thermodynamic pure states.  Recall from
Sec.~\ref{sec:pinning} that if a domain wall, generated by switching from
periodic to antiperiodic boundary conditions, is pinned, then it {\it
must\/} give rise to thermodynamic pure states whose relative interface is
that same domain wall.  So in order for RSB interfaces to be both
space-filling and {\it not\/} give rise to thermodynamic pure states, {\it
they must deflect out of any fixed region\/} as $L\to\infty$.  The
resulting situation is shown in Fig.~3, which would have to occur on all
large lengthscales.

\begin{figure}[t]
\vskip -1 in
\centerline{\epsfig{file=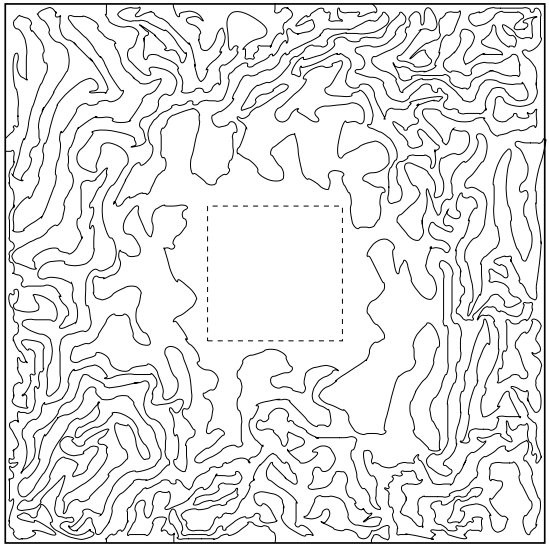,width=3.0in}}
\vskip -0.35in
\renewcommand{\baselinestretch}{1.0} 
\small
\caption{A sketch of RSB interface deflection to infinity, in the situation
where a bulk perturbation is applied to a volume with periodic boundary
conditions at zero temperature, as described in the text. (Two dimensions
is shown only for illustrative simplicity, and is not meant to imply that
mean-field RSB theory is expected to apply there.)  In this figure a
single, positive-density interface is depicted.  Whether the interface
consists of a single or many domain walls is irrelevant, so long as their
union has density that scales as $L^d$.}
\label{fig:rsb}
\end{figure}
\renewcommand{\baselinestretch}{1.25}
\normalsize

It seems already clear that such a situation is highly improbable; in
the next section we prove that, indeed, it cannot occur.

\section{Theorem and proof}
\label{sec:proof}

Consider again the pair of spin configurations $\sigma^L$ and ${\sigma'}^L$
discussed in Sec.~\ref{sec:links}.  We will use the {\it uniform
perturbation metastate\/} introduced in
[NS01]$^{\footnotesize{\cite{NSregcong}}}$.  Here one does for the {\it
pair\/} $(\pm\sigma^L,\pm{\sigma'}^L)$ what was done for $\pm\sigma^L$ in
the original metastate.  The resulting metastate gives, among other things,
a translation-invariant $({\cal J},{\tilde{\cal D}}_{\cal J})$, where
${\tilde{\cal D}}_{\cal J}$ is a domain wall measure that provides the ($L
\to \infty$) probability that a given bond belongs to the relative domain
wall between $\pm\sigma^L$ and $\pm{\sigma'}^L$ inside an arbitrary large
cube $\Lambda_L$.

Now let $(\pm\sigma,\pm\sigma')$ be chosen randomly from the $T=0$
metastate described in the last paragraph.  The argument in [NS01] shows
that, for almost every $({\cal J},\pm\sigma,\pm\sigma')$, either
$\pm\sigma'=\pm\sigma$ or else the two infinite-volume ground state pairs
have a relative interface of strictly positive density (i.e., $\sigma$ and
$\sigma'$ are {\it incongruent\/}, in the terminology of Huse and
Fisher$^{\footnotesize{\cite{HF87a}}}$).  We therefore already know that if
there is a pinned interface at all, it must have strictly positive density,
i.e., $d_s=d$.  We now prove the converse as well.

That is, we prove that if such a space-filling interface exists, then the
situation depicted in Fig.~\ref{fig:rsb} {\it cannot\/} happen.  Such
interfaces must be {\it pinned\/}, i.e., they have {\it strictly
positive\/} probabilities of remaining inside any large fixed window
$\Lambda_{L_0}$ as the outer cube size $L\to\infty$.  Moreover, the
fraction of bonds in the domain wall that remain {\it inside} the window
scales as $(L_0/L)^d$.

{\it Theorem.\/} On each cube $\Lambda_L$, consider
torus-translation-invariant $({\cal J}^L,{\tilde{\cal D}}_{{\cal J}^L})$, a
sequence of random couplings and domain wall measures (from, e.g., the
triple $({\cal J}^L,\pm\sigma^L,\pm{\sigma'}^L)$).  Let $({\cal
J},{\tilde{\cal D}}_{\cal J})$ be any limit in distribution as $L\to\infty$
of $({\cal J}^L,{\tilde{\cal D}}_{{\cal J}^L})$. Then if the probability
that a particular edge belongs to a domain wall is bounded away from zero
as $L\to\infty$, there must be at least a positive fraction of the ergodic
components of $({\cal J},{\tilde{\cal D}}_{\cal J})$ that have a positive
density of domain walls.

{\it Proof.\/} Because the joint distribution of $({\cal J}^L,{\tilde{\cal
D}}_{{\cal J}^L})$ is, for every $L$, invariant under torus translations,
any limiting distribution $({\cal J},{\tilde{\cal D}}_{\cal J})$ is
invariant under all translations of the infinite-volume cubic lattice ${\bf
Z}^d$.  The translation-invariance of $({\cal J},{\tilde{\cal D}}_{\cal
J})$ allows its decomposition into components in which
translation-ergodicity holds (see, e.g.,
[NS96a,NS97b]$^{\footnotesize{\cite{NS96a,NS97}}}$).  For each bond
$\langle x,y\rangle$ consider the event $A_{\langle x,y\rangle}$ that
$\langle x,y\rangle$ is in the domain wall.  If the probability, with
respect to $({\cal J}^L,{\tilde{\cal D}}_{{\cal J}^L})$, of the event
$A_{\langle x,y\rangle}$ occurring in $\Lambda_L$ is larger than some fixed
$\rho>0$ independent of $L$, then any limiting measure must also have the
probability that $A_{\langle x,y\rangle}$ occurs being strictly positive
(and greater than $\rho$).  Because the translation-invariant measure
$({\cal J},{\tilde{\cal D}}_{\cal J})$ therefore has $A_{\langle
x,y\rangle}$ occurring with a probability $P({A_{\langle
x,y\rangle}})>\rho>0$, it follows that in a positive fraction of its
ergodic components, the probability of $A_{\langle x,y\rangle}$ occurring
is also strictly greater than zero.  In each of these ergodic components,
by the spatial ergodic theorem (see, e.g., [NS96a,NS97b]) the spatial
density of $\langle x,y\rangle$'s such that $A_{\langle x,y\rangle}$ occurs
must equal a strictly positive number, i.e., the interface has a nonzero
density.

{\it Remark 1.\/} Although the theorem as formulated here addresses domain
walls between ground states, it should be extendable to domain walls
between spin configurations chosen from different pure states at low
temperature, by ``pruning'' small thermally induced
droplets$^{\footnotesize{\cite{NSregcong}}}$.

{\it Remark 2.\/} Note that the third property of RSB interfaces, namely
that their energy remains of $O(1)$ independently of $L$, was not needed;
the theorem applies to {\it any\/} space-filling domain wall constructed as
discussed. The theorem therefore applies to RSB excitations as a special
case, but to other kinds as well.  These will be discussed further in
Sec.~\ref{sec:discussion}.

We now apply the theorem to RSB interfaces.  These satisfy the condition
that the probability that an arbitrarily chosen bond belongs to a
$(\pm\sigma^L,\pm{\sigma'}^L)$ domain wall is bounded away from zero as
$L\to\infty$.  It therefore follows that, if RSB interfaces exist, and if
one chooses a random {\it infinite-volume\/} GSP
$(\pm\sigma,\pm\sigma')$ from the metastate, then there must be a
positive probability that any given bond belongs to that interface.  This
is equivalent to the statement that, for every $L$ as $L\to\infty$, there
is a positive probability of finding an RSB interface inside any fixed
window $\Lambda_{L_0}$ of arbitrary (but finite) size $L_0$.

Moreover, if pinned RSB interfaces are present at $T=0$, then they
would presumably give rise to multiple pure state pairs at low but nonzero
temperature.  Here it would be the case that, in addition to the expected
thermal fluctuations, two spin configurations, each randomly chosen from
different pure states (not globally flip-related), would have a relative
RSB interface.  Equivalently, one could determine the existence of these
positive-temperature interfaces by examining the thermal expectations of,
e.g., two-point correlation functions.

It follows that the mean-field RSB theory {\it must give rise to multiple
thermodynamic ground state pairs at $T=0$, and by extension, pure
state pairs as conventionally defined (cf.~Appendix~A) also at low $T$.\/}

\section{Discussion and conclusion}
\label{sec:discussion}

We have shown that the claim in [MPRRZ]$^{\footnotesize{\cite{MPRRZ}}}$
(elaborated on in Secs.~3 and 7 of that paper) that mean-field RSB theory
does {\it not\/} give rise to the usual thermodynamic pure states is in
rigorous contradiction with the simultaneous claim that the theory also
predicts interfaces (equivalently, nontrivial link overlap distribution
functions) with the properties delineated in Sec.~\ref{sec:links}.  Then,
in short-ranged spin glasses in finite dimensions, either there are {\it
no\/} interfaces that are both space-filling, with $d_s=d$, and have energy
of $O(1)$, or else there are and they comprise domain walls between
distinct thermodynamic pure state pairs.  We investigate each of these
possibilities in turn.

The first possibility is that there are no domain walls that are both
space-filling and have $O(1)$ energy.  Suppose we relax the second
requirement, so that the energy of the interface increases with $L$.  The
simplest possibility is that this energy scales as $L^{\theta'}$ with
$\theta'>0$, although more slowly increasing functions are also possible
(e.g., $\log(L)$); but because the argument is the same for both, we simply
examine the case $\theta'>0$.  But when
$\theta'>0$, one recovers$^{\footnotesize{\cite{NS98}}}$ the chaotic pairs
picture, as noted in Sec.~\ref{sec:links}; multiple pure state pairs cannot
now coexist at any $T$ within a single $\Lambda_L$ with large $L$. 

Relaxing the first requirement implies that there are either no domain
walls at all, in which case one recovers a two-state picture, or else there
are interfaces with $d_s<d$.  The latter would not be seen in any $T=0$
metastate constructed using coupling-independent boundary
conditions$^{\footnotesize{\cite{NS2D00,NS2Dlong}}}$, and may or may not give rise
to new pure states at $T>0$ depending on how they are constructed; an
extensive discussion is given in
[NS01]$^{\footnotesize{\cite{NSregcong}}}$.  In either case any metastate
constructed using coupling-independent boundary conditions would see at
most only a single pair of pure states (cf.~the discussion of ``invisible''
states within the metastate given in Sec.~V of
[NS98]$^{\footnotesize{\cite{NS98}}}$).

We turn now to the second possibility in which space-filling domain walls
of $O(1)$ energy {\it are\/} present on all lengthscales.  Now the theorem
in Sec.~\ref{sec:proof} necessitates the existence of multiple pure
state pairs so that the thermodynamic states $\Gamma$
(cf.~Sec.~\ref{subsec:metastates}) would be mixed states at $T>0$.  

The conclusion is that regardless of any re-interpretations of the
``meaning'' of the RSB ansatz for finite-dimensional spin glasses, it has
an unambiguous prediction for the structural difference in ground states
generated in a large cube when a KMPY-type perturbation ([KM,
PY]$^{\footnotesize{\cite{KM00,PY00}}}$), as examined in
[MP00b]$^{\footnotesize{\cite{MP00b}}}$, is applied.  This prediction is
the presence of space-filling interfaces.  But we have shown here that this
feature gives rise to multiple {\it infinite-volume\/} pure states.  So the
RSB ansatz predicts these multiple thermodynamic states whether it was
originally intended to or not.

Moreover, given that these RSB interfaces have energies of $O(1)$, the
thermodynamic states they give rise to would necessarily appear as mixed
states in large finite volumes, as in Eq.~(\ref{eq:sumfinite}). But this
possibility was ruled out in our earlier
papers$^{\footnotesize{\cite{NS96a,NS96b,NSBerlin,NS97,NS98}}}$ (and was
disavowed in [MPRRZ]).

Mean-field RSB theory therefore can not describe the low-temperature
structure of the spin glass phase in any {\it finite\/} dimension, although
of course RSB theory presumably remains valid for infinite-ranged models.
If the low-temperature spin glass phase displays multiple pure states in
any finite dimension, their structure would have to be given by the chaotic
pairs picture of
[NS96b,NS97,NS98]$^{\footnotesize{\cite{NS96b,NS97,NS98}}}$ and spin
overlap structures inside any window would be trivial regardless of how the
overlaps are constructed.

\medskip

{\it Acknowledgments.}  We are grateful to Peter Young for useful
correspondence, and to Michael E.~Fisher for numerous helpful comments and
suggestions on the manuscript. We also thank David Huse and Joel Lebowitz
for their comments.

\renewcommand{\baselinestretch}{1.0}

\renewcommand{\baselinestretch}{1.25}
\normalsize

\appendix
\section{Gibbs states in finite and infinite volume}
\renewcommand{\theequation}{{\Alph{section}.\arabic{equation}}}
\setcounter{equation}{0}
\label{app:A}

In this appendix, we present some background information about Gibbs
states, closely following the discussion in
[NS97]$^{\footnotesize{\cite{NS97}}}$.  Given the EA Hamiltonian
(\ref{eq:Hamiltonian}) on $\Lambda_L$ with a specified boundary condition
(e.g., free, fixed, periodic, etc.), the finite-volume Gibbs state
$\rho_{{\cal J},T}^{(L)}$ on $\Lambda_L$ at temperature $T$ is defined by:
\begin{equation}
\label{eq:finite}
\rho_{{\cal J},T}^{(L)}(\sigma)=Z_{L,T}^{-1} \exp \{-{\cal H}_{{\cal
J},L}(\sigma)/k_BT\}\quad ,
\end{equation}
where the partition function $Z_{L,T}$ is such that the sum of
$\rho_{{\cal J},T}^{(L)}$ over all spin configurations in $\Lambda_L$
yields one.

The finite-volume Gibbs state $\rho_{{\cal J},T}^{(L)}(\sigma)$ is a
probability measure, describing at fixed $T$ the likelihood of a given spin
configuration obeying the specified boundary condition appearing within
$\Lambda_L$.  Equivalently, the measure is specified by the set of all
correlation functions within $\Lambda_L$, i.e., by the set of all
$\langle\sigma_{x_1}\cdots\sigma_{x_m}\rangle$ for arbitrary $m$ and
arbitrary $x_1,\ldots,x_m\in\Lambda_L$.

A {\it thermodynamic\/} state $\rho_{{\cal J},T}$ is defined as an {\it
infinite\/}-volume Gibbs measure, containing information such as the
probability of any finite subset of spins taking on specified values.
Thermodynamic states can be constructed by taking the $L\to\infty$ limit of
a sequence of finite-volume Gibbs states $\rho_{{\cal J},T}^{(L)}(\sigma)$,
each with a specified boundary condition (which may remain the same or may
change with $L$).  The idea of a limiting measure can be made precise by
requiring that every $m$-spin correlation function, for $m=1,2,\ldots$,
possesses a limit.  Infinite-volume Gibbs measures $\rho_{{\cal J},T}$ can
also be characterized independently of any limiting process, as probability
measures on infinite-volume spin configurations that satisfy the
Dobrushin-Lanford-Ruelle (DLR) equations (for a mathematically detailed
presentation, see the book of Georgii$^{\footnotesize{\cite{Georgii}}}$).

Thermodynamic states may or may not be mixtures of other thermodynamic
states.  If a Gibbs state $\rho_{{\cal J},T}$ can be decomposed
according to
\begin{equation}
\label{eq:mixed}
\rho_{{\cal J},T}=\lambda\rho^{1}_{{\cal J},T}+(1-\lambda)\rho^{2}_{{\cal J},T}\quad ,
\end{equation}
where $0 < \lambda < 1$ and $\rho^{1}$ and $\rho^{2}$ are also
infinite-volume Gibbs states (distinct from $\rho$), then $\rho_{{\cal
J},T}$ is a {\it mixed\/} thermodynamic state or simply, mixed state.  A
mixed state may have as few as two or as many as an uncountable infinity of
states in its decomposition.  The meaning of Eq.~(\ref{eq:mixed}) can be
understood as follows: any correlation function computed using the
thermodynamic state $\rho_{{\cal J},T}$ can be decomposed in the following
way:
\begin{equation}
\label{eq:decomposed}
\langle\sigma_{x_1}\cdots\sigma_{x_m}\rangle_{\rho_{{\cal J},T}} = 
\lambda\langle\sigma_{x_1}\cdots\sigma_{x_m}\rangle_{\rho^{1}_{{\cal
J},T}}
+(1-\lambda)\langle\sigma_{x_1}\cdots\sigma_{x_m}
\rangle_{\rho^{2}_{{\cal J},T}}\quad .
\end{equation}

If a state cannot be written as a convex combination of any other
infinite-volume Gibbs states, it is then a thermodynamic {\it pure\/}
state.  As an illustration, the paramagnetic state is a pure state, as are
each of the positive and negative magnetization states in the Ising
ferromagnet.  In that same system, the Gibbs state produced by a sequence
of increasing volumes, at $T<T_c$, using only periodic or free boundary
conditions is a mixed state, decomposable into the positive and negative
magnetization states, with $\lambda=1/2$.  A thermodynamic pure state
$\rho_P$ can be intrinsically characterized by a {\it clustering
property\/} (see, e.g., Refs.~\cite{Georgii,vEvH}), which implies that for
any fixed $x$,
\begin{equation}
\label{eq:clustering}
\langle\sigma_x\sigma_y\rangle_{\rho_P} - 
\langle\sigma_x\rangle_{\rho_P}\langle\sigma_y\rangle_{\rho_P}\ \to 0 ,
\qquad \vert y \vert\to\infty\quad ,
\end{equation}
and similar clustering for higher order correlations.

Finite-volume Gibbs states, which are well-defined probability measures,
should not be confused with the putative ``finite-volume pure states'' of
[MPRRZ]$^{\footnotesize{\cite{MPRRZ}}}$, which have not been clearly
defined.  A finite-volume Gibbs state can have an {\it approximate\/}
decomposition into {\it thermodynamic\/} pure states restricted to a
``window''$^{\footnotesize{\cite{NS98}}}$ deep inside $\Lambda_L$, as in
Eq.~(\ref{eq:sum}).  Whether a similar decomposition of finite-volume Gibbs
states into ``finite-volume pure states'' can be made is unclear; it would
at the least require making the notion of finite-volume pure state more
precise.

\section{Glossary}
\renewcommand{\theequation}{{\Bet{section}.\arabic{equation}}}
\setcounter{equation}{0}
\label{app:B}

We include this glossary for the reader's convenience.  All definitions
here are informal.  Terms that have appeared only recently in the
literature, or that may be less familiar, are also defined within the text;
in such cases, the section where they are first defined is also noted.

\bigskip

\noindent {\bf Chaotic Pairs Picture} A scenario for the low-temperature spin glass
phase in which there exist infinitely many (incongruent) pure state pairs
(for a.e.~${\cal J}$) at all temperatures below $T_c$, but with probability
one only a single one of these pairs would be seen in any large volume with
periodic boundary conditions.  The overlap function computed in any large
volume is therefore indistinguishable from a two-state picture like
droplet/scaling (cf.~Fig.~\ref{fig:overlap}).  However, the pair chosen
varies chaotically with volume.  (Sec.~\ref{subsec:metastates}.)

\medskip

\noindent {\bf Chaotic Size Dependence} Inside any large volume $\Lambda_L$ with
specified boundary conditions, the Gibbs state is approximately either a
single pure state (e.g., in a homogeneous Ising ferromagnet, one has a
paramagnet above $T_c$ for any boundary condition, the magnetized plus
state below $T_c$ for all plus spins at the boundary, etc.), or else an
approximate decomposition over pure states as in Eq.~\ref{eq:sumfinite}
(e.g., in the same system below $T_c$ the Gibbs state is an equal mixture
of the magnetized plus and minus states).  Chaotic size dependence occurs
when the pure states and/or weights vary persistently as $L$ is increased,
so that there is no limiting infinite-volume Gibbs state.
(Secs.~\ref{subsec:RSB} and \ref{subsec:metastates}.)

\medskip

\noindent {\bf Deflection to Infinity} Consider an interface between two
ground or pure states in $\Lambda_L$ generated either by a change in
boundary condition (e.g., periodic to antiperiodic), or by addition of a
perturbation with a single boundary condition.  Consider a volume
$\Lambda_{L_0}$ of arbitrary but fixed side $L_0$.  If, for any $L_0$, the
relative interface eventually moves (and stays) outside of $\Lambda_{L_0}$
as $L\to\infty$, the interface has ``deflected to infinity''.  See
Fig.~\ref{fig:deflect} for an illustration.  (Sec.~\ref{sec:pinning}.)

\medskip

\noindent {\bf Droplet/Scaling} A two-state picture (see below) whose
properties follow from a scaling {\it ansatz\/} developed by
Macmillan$^{\footnotesize{\cite{Mac}}}$, Bray and
Moore$^{\footnotesize{\cite{BM85,BM87}}}$, and Fisher and
Huse$^{\footnotesize{\cite{FH86,HF87a,FH87b,FH88}}}$; the last of these
fully developed the physical droplet picture corresponding to the scaling
ansatz, which followed from ``domain wall'' renormalization-group studies
of the first two groups.  In this picture, the thermodynamic and dynamic
properties of spin glasses at low temperature are dominated by low-lying
excitations corresponding to clusters of coherently flipped spins.  The
density of states of these clusters at zero energy falls off as a power law
in lengthscale $L$, with exponent bounded from above by $(d-1)/2$.  At low
temperatures and on large lengthscales the density of thermally activated
clusters is dilute and they can be considered as non-interacting two-level
systems.

\medskip

\noindent {\bf Ground State} In a finite volume $\Lambda_L$, the
lowest-energy state(s) consistent with the boundary conditions.  A
convergent sequence of finite-volume ground states yields an
infinite-volume ground state, which is simply a pure state (as in
Appendix~A) at $T=0$.  An infinite-volume ground state can alternatively
(and often more usefully) be defined as an infinite-volume spin
configuration whose energy cannot be lowered by the flip of any finite
subset of spins. (Sec.~\ref{sec:pinning}.)

\medskip

\noindent {\bf Ground State Pair (GSP)} In the absence of an external field
or spin-flip symmetry breaking boundary conditions, ground states occur in
pairs related by a global spin flip.

\medskip
  
\noindent {\bf Incongruence} Two spin configurations are incongruent (a
notion introduced by Huse and Fisher$^{\footnotesize{\cite{HF87a}}}$) if
they differ by a relative flip along a space-filling interface; that is, a
nonzero density of bonds is satisfied in one but not the other spin
configuration.  If the relative interface has zero density, the spin
configurations are said to be {\it regionally congruent\/}.
(Sec.~\ref{sec:proof}.)

\medskip

\noindent {\bf Pinning} Given the same scenario as in the definition of
{\bf deflection to infinity} above, the interface remains inside a
sufficiently large volume of fixed size $L_0$ as
$L\to\infty$. (Sec.~\ref{sec:pinning}.)

\medskip

\noindent {\bf Metastate} A probability measure on infinite-volume
thermodynamic states that carries all relevant thermodynamic information
about a system.  In the current context, the metastate provides, among
other things, the probability of appearance of various pure or ground
states appearing within a large finite volume $\Lambda_L$ with specified
boundary conditions.  (Sec.~\ref{subsec:metastates}.)

\medskip

\noindent {\bf RSB Interface} An interface between two globally different
spin configurations (that are not global flips of each other) that has the
properties of being both space-filling and also of having approximately
order one energy independently of lengthscale.  (Sec.~\ref{sec:links}.)

\medskip

\noindent {\bf Two-State Picture} A scenario for the low-temperature spin
glass phase in which there exists only a single pair of (spin-flip-related)
pure states at all temperatures below $T_c$.  Although not used in the
text, we note that these can be divided into at least two kinds.  A {\it
strong\/} two-state picture is one where there are no more than two pure
states at any temperature, as in the droplet/scaling picture.  A {\it
weak\/} two-state picture is one where there exists a ``special'' pair of
pure states that supports any metastate generated by coupling-independent
boundary conditions, but in which there also exist other pure states that
can be generated only by coupling-dependent boundary conditions.  These
latter states are ``invisible'' in any coupling-independent b.c.~metastate.
(This possibility for spin glasses is very briefly discussed in
Sec.~\ref{sec:discussion}.)  An example of a weak two-state picture could
be the homogeneous Ising ferromagnet, at $T=0$ in two dimensions and below
the roughening temperature in three and higher dimensions.  Here the
special pair is the uniformly magnetized plus and minus states, while the
others are the interface states.

\medskip

\noindent {\bf Ultrametricity} In the spin glass context, the property that
the joint overlap statistics of any three pure states with overlaps $q_1$,
$q_2$, and $q_3$ satisfy the condition $q_1=q_2\le q_3$, consistent with a
hierarchical pure state structure.

\medskip

\noindent {\bf Window} Given a large volume $\Lambda_L$ with specified
boundary conditions, a window is an interior volume $\Lambda_{L_0}$ with
$1\ll L_0\ll L$. Both $\Lambda_L$ and $\Lambda_{L_0}$ are centered at the
origin.  We argue in [NS98]$^{\footnotesize{\cite{NS98}}}$ that an overlap
computation must be done inside a window if it is to reveal any unambiguous
information about pure state structure. (Sec.~\ref{subsec:windows}.)

\end{document}